\documentclass[ALICE,manyauthors]{cernphprep}
\usepackage[comma,square,numbers,sort&compress]{natbib}
\usepackage{caption}
\RequirePackage{booktabs,amsmath}
\usepackage{hyperref}
\usepackage{lineno,multirow}
\newcommand{\jpsi}{\ensuremath{\mathrm{J}/\psi}}
\newcommand{\psip}{\ensuremath{\psi(2S)}}
\newcommand{\bpsip}{$\mathbf{\ensuremath{\psi(2S)}}$}

\newcommand{\pt}{$p_{\rm T}$}
\newcommand{\snn}{$\sqrt{s_\mathrm{NN}}$}
\newcommand{\bsnn}{$\sqrt{s_\mathrm{NN}}$}

\newcommand{\acceps}{\ensuremath{(\mathrm{Acc}\times\varepsilon)_{\psip}}}
\modulolinenumbers[5]

\newcommand{\minvm}{$M_\mathrm{\pi^+\pi^-\mu^+\mu^-}$}
\newcommand{\minve}{$M_\mathrm{\pi^+\pi^-e^+e^-}$}

\begin{document}%

%  Title page %
\begin{titlepage}
\PHyear{2015}
\PHnumber{156}      % required, will be obtained from PH
\PHdate{22 June}  % required, will be obtained from PH
%

% Put your own title + short title here:
\title{Coherent \bpsip~photo-production\\in ultra-peripheral Pb--Pb collisions at $\mathbf{\sqrt{\textit{s}_{\rm NN}}}$ = 2.76 TeV}
\ShortTitle{Coherent \bpsip~photo-production at \bsnn = 2.76 TeV}   % appears on right page headers

% Do not change the next lines
\Collaboration{ALICE Collaboration\thanks{See Appendix~\ref{app:collab} for the list of collaboration members}}
\ShortAuthor{ALICE Collaboration} % appears on left page headers, do not change

\begin{abstract}
We have performed the first measurement of the coherent $\psi$(2S)~photo-production cross section
in ultra-peripheral Pb--Pb collisions at the LHC. This charmonium excited state is reconstructed via the 
$\psi$(2S) $\rightarrow l^{+}l^{-}$ and 
$\psi$(2S) $\rightarrow$ J/$\psi \pi^{+}\pi^{-}$~decays, where the J/$\psi$~decays into two leptons. 
The analysis is based on an event sample corresponding to an integrated 
luminosity of about 22 $\mu\rm{b}^{-1}$. The cross section for coherent $\psi$(2S)~production in the 
rapidity interval $-0.9<y<0.9$ is 
$\mathrm{d}\sigma_{\psi{\rm(2S)}}^{\rm coh}/\mathrm{d}y =0.83\pm 0.19\big(\mathrm{\rm{stat}+{\rm syst}}\big)$~mb. 
The $\psi$(2S) to J/$\psi$~coherent cross section ratio is $0.34^{+0.08}_{-0.07}(\rm{stat}+{\rm syst})$. 
The obtained results are compared to predictions from theoretical models.

\end{abstract}
\end{titlepage}
\setcounter{page}{2}
%\begin{keyword}
%Heavy ions, Vector mesons, Photo-production, ALICE, LHC.
%\end{keyword}

%\end{frontmatter}

%\linenumbers

\section{Introduction}
Two-photon and photo-nuclear interactions at unprecedented energies can be studied in heavy-ion Ultra-Peripheral 
Collisions (UPC) at the LHC. In such collisions the nuclei are separated
by impact parameters larger than the sum of their radii and therefore hadronic interactions are
strongly suppressed. The cross sections for photon induced reactions remain large because
the strong electromagnetic field of the nucleus enhances the intensity of the photon flux,
which grows as 
the square of the charge
of the nucleus. The physics of ultra-peripheral
collisions is reviewed in~\cite{Baltz:2007kq,Bertulani:2005ru}. Exclusive photo-production of vector mesons at high energy, where a vector
meson is produced in an event with no other final state particles, is of particular interest, since
it provides a measure of the nuclear gluon distribution at low Bjorken-$x$.

Exclusive production of charmonium in photon-proton interactions at HERA~\cite{Chekanov:2002xi,Alexa:2013xxa,Adloff:2002re}, 
$\gamma +p\rightarrow\newline\jpsi~(\psip)+p$, 
has been successfully modelled 
in terms of the exchange
of two gluons with no net-colour transfer ~\cite{Ryskin:1992ui}. Experimental data on this process from HERA
have been used to constrain the proton gluon distribution at low Bjorken-$x$~\cite{Martin:2007sb}. Exclusive
vector meson production in heavy-ion interactions is expected to probe the nuclear gluon distribution~\cite{Rebyakova:2011vf}, 
for which there is considerable uncertainty in the low-$x$ region~\cite{Eskola:2009uj}. 
Exclusive $\rho^{0}$~\cite{Abelev:2007nb} and \jpsi~\cite{Afanasiev:2009hy} production has been studied in Au-Au collisions at RHIC. 
The exclusive
photo-production can be either coherent, where the photon couples coherently to almost all the
nucleons, or incoherent, where the photon couples to a single nucleon. Coherent production is
characterized by low transverse momenta of vector mesons~(\pt~$\simeq$ 60 MeV/$c$) where the
target nucleus normally does not break up. However, the exchange of additional photons, radiated 
independently from the original one, may
lead to the target nucleus breaking up or de-excite through neutron emission.
Simulation models estimate this occurs in about 30$\%$ of
the events~\cite{Baltz:2002pp}. Incoherent production is
characterized by a somewhat higher transverse momentum of the vector mesons~(\pt~$\simeq$~500 MeV/$c$). In
this case the nucleus interacting with the photon breaks up but, apart from single nucleons or
nuclear fragments in the very forward region, no other particles are produced besides the vector meson.

We published the first results on the coherent photo-production of \jpsi~in UPC
 Pb--Pb collisions at the LHC~\cite{Abelev:2012ba} in the rapidity region \mbox{--3.6~$< y <$--2.6}, which constrain 
the nuclear gluon distribution at Bjorken-$x$~$\simeq 10^{-2}$. Shortly afterwards, ALICE
published a second paper measuring both the coherent and the incoherent \jpsi~vector meson cross section at mid-rapidity~\cite{Abbas:2013oua},
allowing the nuclear gluon distribution at Bjorken-$x$~$\simeq 10^{-3}$ to be explored. The present analysis is performed in the 
same rapidity region with respect to the measurement reported in ~\cite{Abbas:2013oua}, and it is sensitive 
to Bjorken-$x$~$\simeq 10^{-3}$ too.    

There are very few studies of photo-production of \psip~off nuclei.
Incoherent photo-production, using a 21~GeV photon beam off a deuterium target, has been studied in~\cite{Camerini:1975cy};
non-exclusive photo-production, using bremstrahlung photons with an average energy of 90 GeV off a $^{6}$Li target, have been
reported in~\cite{Barate:1986fq}. However, no previous measurements of \psip~coherent photo-production off
nuclear targets, have been reported in the literature. 

In this letter, results from ALICE on exclusive coherent photo-production of \psip~mesons 
at mid-rapidity in ultra-peripheral Pb--Pb collisions at 
$\sqrt{s_{\mathrm{NN}} } = 2.76$ TeV are presented. 
The measured coherent \psip~cross section and 
the \psip/\jpsi~cross section ratio
are compared to model predictions~\cite{Klein:1999qj,Adeluyi:2013tuu,Ducati:2013bya,Goncalves:2011vf,Lappi:2013am,Guzey:2014kka}.

\section{Detector description} 
The main components of the ALICE detector are a central barrel placed in a large solenoid magnet (B = 0.5 T),
covering the pseudo-rapidity region $\vert\eta\vert$ $<$ 0.9, and a muon
spectrometer at forward rapidity, covering the range 
\hbox{--4.0~$<\eta<$~--2.5~\cite{Aamodt:2008zz}}.
Three central barrel detectors are used in this analysis. 
The ALICE Internal Tracking System(ITS) is made of six silicon layers, all of them used in this analysis for particle tracking. 
The Silicon Pixel Detector (SPD) makes up the two innermost layers of the ITS, covering  
pseudo-rapidity ranges $|\eta| < 2$ and $|\eta| < 1.4$, for the inner (radius 3.9 cm)  and outer (average radius 7.6 cm) layers, 
respectively. 
The SPD is a fine granularity detector, having about $10^{7}$ pixels, and is used for triggering purposes. 
The Time Projection Chamber (TPC) is used for tracking and for particle identification~\cite{Alme:2010ke} and has
an acceptance covering the pseudo-rapidity region $|\eta| < 0.9$. 
The Time-of-Flight detector (TOF) surrounds the TPC and is 
a large cylindrical barrel of Multigap Resistive Plate Chambers (MRPC) with about 150,000 readout channels, giving very high precision timing. 
The TOF pseudo-rapidity coverage matches that of the TPC. Used in combination with the tracking system, the TOF detector is used for charged particle 
identification up to a transverse momentum of about 2.5 GeV/$c$ (pions and kaons) and 4 GeV/$c$ (protons). 
In addition, the TOF detector is also used for triggering~\cite{Akindinov:2009zzc}.
 
The analysis presented below also makes use of two forward detectors.
 The V0 counters consist of 
two arrays of 32 scintillator tiles each, covering the range 2.8~$<$ $\eta$ $<$~5.1
(V0-A, on the opposite side of the muon spectrometer) and --3.7~$<$~$\eta$~$<$~--1.7
(V0-C, on the same side as the muon spectrometer) and positioned respectively at $z$~= 340~cm and $z$~= --90~cm
from the interaction point. 

Finally, 
two sets of hadronic Zero Degree Calorimeters (ZDC) are located at 114~m on either side 
of the interaction point. The ZDCs detect neutrons emitted in the very forward and backward regions~($|\eta|>8.7$), such as 
neutrons produced by electromagnetic dissociation of the nucleus~\cite{ALICE:2012aa}~(see Section 3).
\section{Data Analysis}
The event sample considered for the present analysis was collected during the 2011 Pb--Pb run, using a
dedicated Barrel Ultra-Peripheral Collision trigger (BUPC), 
selecting events with  
the following characteristics: \newline
(i) at least two hits in the SPD detector; \\
(ii) a number of fired pad-OR ($N^{on}$) in the TOF detector~\cite{Akindinov:2009zzc} in the range 2~$\leq N^{on}$$\leq$~6, 
with at least two of them with a difference in azimuth,  $\Delta \phi$, in the range $150^\circ \leq \Delta \phi \leq $180$^\circ$; \\
(iii) no hits in the V0-A and no hits in the~\hbox{V0-C} detectors. \\
The integrated luminosity used in this analysis was 22.4~$^{+0.9}_{-1.2}~\mu\rm{b}^{-1}$. 
Luminosity determination and systematics are discussed in Sec. 3.1
In the present analysis, coherent \psip~photo-production was studied in four different channels: 
\psip $\rightarrow l^+l^-$
and 
\psip $\rightarrow \jpsi~\pi^{+}\pi^{-}$, followed by the 
\jpsi $\rightarrow l^{+}l^{-}$ decay, where $l^{+}l^{-}$ can be 
either a $e^{+}e^{-}$ or $\mu^{+}\mu^{-}$ pair.
%
%***************** section 3.1 ***************
%
\subsection{\texorpdfstring{The $\psip\rightarrow l^+l^-$ channel}{}}
For the di-muon and di-electron decay channels
the following selection criteria were applied: \newline
(i) a reconstructed primary vertex. The primary vertex position is determined from the tracks reconstructed
in the ITS and TPC as described in ref.~\cite{vertex:2012}. The vertex reconstruction algorithm is fully
efficient for events with at least one reconstructed primary charged particle in the common 
TPC and ITS acceptance;\\
(ii) only two good tracks with at least 70 TPC clusters and at least 1 SPD cluster each.
Moreover, particles originated in secondary hadronic interactions or conversions in the detector
material, were removed using a distance of closest approach ~($\rm{DCA}$) cut. 
The tracks extrapolated to the reconstructed
vertex should have a $DCA$ in the 
beam direction $DCA_{L}\leq$~2~cm, and 
in the plane orthogonal to the beam direction
$DCA_{T}\leq$~0.0182~+~0.0350/$p_{\rm T}^{1.01}$, where $p_{\rm T}$ is the transverse momentum in~(GeV/$c$)~\cite{Alice:900GeV};\\
(iii) at least one of the two good tracks selected in criterion~(ii) should have \pt~$\geq$ 1 GeV/$c$; this cut 
reduces the background, while 
it marginally affects the genuine leptons from \jpsi~decays;\\ 
(iv) The V0 trigger required no signal within a time window of 25~ns around the collision time 
in any of the scintillator tiles of both V0-A and V0-C. 
Signals in both V0 detectors were searched offline in a larger window according to the prescription described in~\cite{Abbas:2013oua};\\
(v) the specific energy loss d$E$/d$x$ for the two tracks is compatible with that of electrons or muons~(see below); 
it is worth noting that the TPC 
resolution does not allow muon and charged pion discrimination;\\
(vi) the two tracks have opposite charges. \\

\begin{table}[htb!p]
\centering
\begin{tabular}{cccc}
\toprule
 & $\rm \psi(2S) \rightarrow l^{+} l^{-}$ & $\rm \psi(2S) \rightarrow \mu^{+} \mu^{-} \pi^{+} \pi^{-}$ & $\rm \psi(2S) \rightarrow e^{+} e^{-} \pi^{+} \pi^{-}$ \\
\midrule
Signal counts & 18.4 $\pm$ 9.3 & 17 $\pm$ 4.1 & 11.0 $\pm$ 3.3 \\
Bkg. counts$(N^{back})$  & 0 & 1 & 0\\
$f_{\rm I}$ & $(5.6 \pm 1.8)\%$ & $(3.4\pm 1.1)\%$ & $(13.2 \pm 4.3)\%$\\
$(\rm Acc\times \epsilon)_{\psi(2S)}$ & $(3.65\pm 0.16)\%$& $(2.35\pm 0.14)\%$& $(1.33\pm 0.08)\%$\\
BR & $(1.56\pm 0.11)\%$ & $(2.02\pm0.03)\%$ & $(2.02\pm 0.03)\%$\\
$\mathcal{L}_{\rm int}$ &(22.4~$^{+0.9}_{-1.2})~\mu\rm{b}^{-1}$ & (22.4~$^{+0.9}_{-1.2})~\mu\rm{b}^{-1}$& (22.4~$^{+0.9}_{-1.2})~\mu\rm{b}^{-1}$\\
$\Delta y$ & 1.8 & 1.8& 1.8\\
\addlinespace
\midrule
$\frac{\rm {d}\sigma^{\rm coh}_{\psi(2S)}}{\rm{d} y}$ (mb)& 0.76 $\pm$ 0.40(stat) $\pm$ 0.13(syst) & 0.81 $\pm$ 0.22(stat) $^{+0.09}_{-0.10}$(syst) & 0.90 $\pm$ 0.31(stat) $^{+0.13}_{-0.12}$(syst)\\
\bottomrule
\end{tabular}
\captionsetup{justification=raggedright,singlelinecheck=false}
\caption{Summary of the main experimental results and correction parameters used in the cross section evaluation. The bottom line shows the cross section 
for the three \psip~decay channels.}
\label{tab:CrossSection}
\end{table}

\begin{table}[htb!p]
\centering
\begin{tabular}{ll ccc}
\toprule
 & & $\rm \psi(2S) \rightarrow l^{+} l^{-}$ & $\rm \psi(2S) \rightarrow \mu^{+} \mu^{-} \pi^{+} \pi^{-}$ & $\rm \psi(2S) \rightarrow e^{+} e^{-} \pi^{+} \pi^{-}$  \\
\midrule
\multicolumn{2}{l}{Signal extraction} &$\pm$ 12\% & $<$ 1\% & $<$ 1\% \\
\addlinespace
\multicolumn{2}{l}{Incoherent contamination~($f_I$)} &$\pm$ 1.8\% &$\pm$ 1.3\% &$\pm$ 4.8\%  \\
\addlinespace
\multirow{2}{*}{$(\rm Acc\times \epsilon)$} & Generator $\frac{\rm {d}\sigma}{\rm{d} y}$ &$\pm$ 1\% &$\pm$ 2\% &$\pm$ 2\% \\
& Tracking efficiency & $\pm$ 4.2\% & $\pm$ 6.0\% & $\pm$ 6.0\% \\
\addlinespace
\multicolumn{2}{l}{Trigger efficiency} & $^{+4\%}_{-9\%}$ & $^{+4\%}_{-9\%}$ & $^{+4\%}_{-9\%}$ \\
\addlinespace
\multicolumn{2}{l}{$e/\mu$ separation} &$\pm$ 2\% &$\pm$ 2\% &$\pm$ 2\% \\
\addlinespace
\multicolumn{2}{l}{V0 offline decision} & $^{+6\%}_{-0\%}$ & $^{+6\%}_{-0\%}$ & $^{+9\%}_{-0\%}$ \\
\addlinespace
\multicolumn{2}{l}{Luminosity} & $^{+5.5\%}_{-4.0\%}$ & $^{+5.5\%}_{-4.0\%}$ & $^{+5.5\%}_{-4.0\%}$ \\
\addlinespace
\multicolumn{2}{l}{Branching ratio} &$\pm$ 7.1\% &$\pm$ 1.5\% &$\pm$ 1.5\% \\
\addlinespace
\midrule
\multicolumn{2}{l}{Uncorrelated sources} &$\pm$ 13\%  & $\pm$ 2\% & $\pm$ 5\% \\
\addlinespace
\multicolumn{2}{l}{Correlated sources} & $^{+10\%}_{-11\%}$ & $^{+11\%}_{-12\%}$ & $^{+13\%}_{-12\%}$\\
\addlinespace
\multicolumn{2}{l}{Total} & $\pm$ 17\% & $^{+11\%}_{-12\%}$ & $^{+14\%}_{-13\%}$ \\
\bottomrule
\end{tabular}
\captionsetup{justification=raggedright,singlelinecheck=false}
\caption{Systematic uncertainties per decay channel.}
\label{tab:Systematics}
\end{table}

\begin{table}[htb!p]
\vskip 0.4 cm
\centering
\begin{tabular}{c c c c c}
\toprule
 & Data & Fraction &STARLIGHT & RSZ \\ 
\midrule
\addlinespace
0n~0n & 20 & (71 $^{+9}_{-11}) \% $& 66\% & 70\% \\
\addlinespace
Xn & 8 & (29 $^{+11}_{-9}) \% $& 34\% & 30\% \\
\addlinespace
0n~Xn & 7 & (25 $^{+11}_{-9}) \% $& 25\% & 23\% \\
\addlinespace
Xn~Xn & 1 & (4 $^{+8}_{-3})\% $& 9\% & 7\% \\
\addlinespace
\bottomrule
\end{tabular}
\captionsetup{justification=raggedright,singlelinecheck=false}
\caption{Number of events for different neutron emissions in the 
$\rm \psi(2S) \rightarrow l^{+} l^{-} \pi^{+} \pi^{-}$ process.}
\label{tab:NeutronZDC}
\end{table}

\begin{figure}[tbh!p]
\begin{center}
\includegraphics[width=7cm,keepaspectratio]{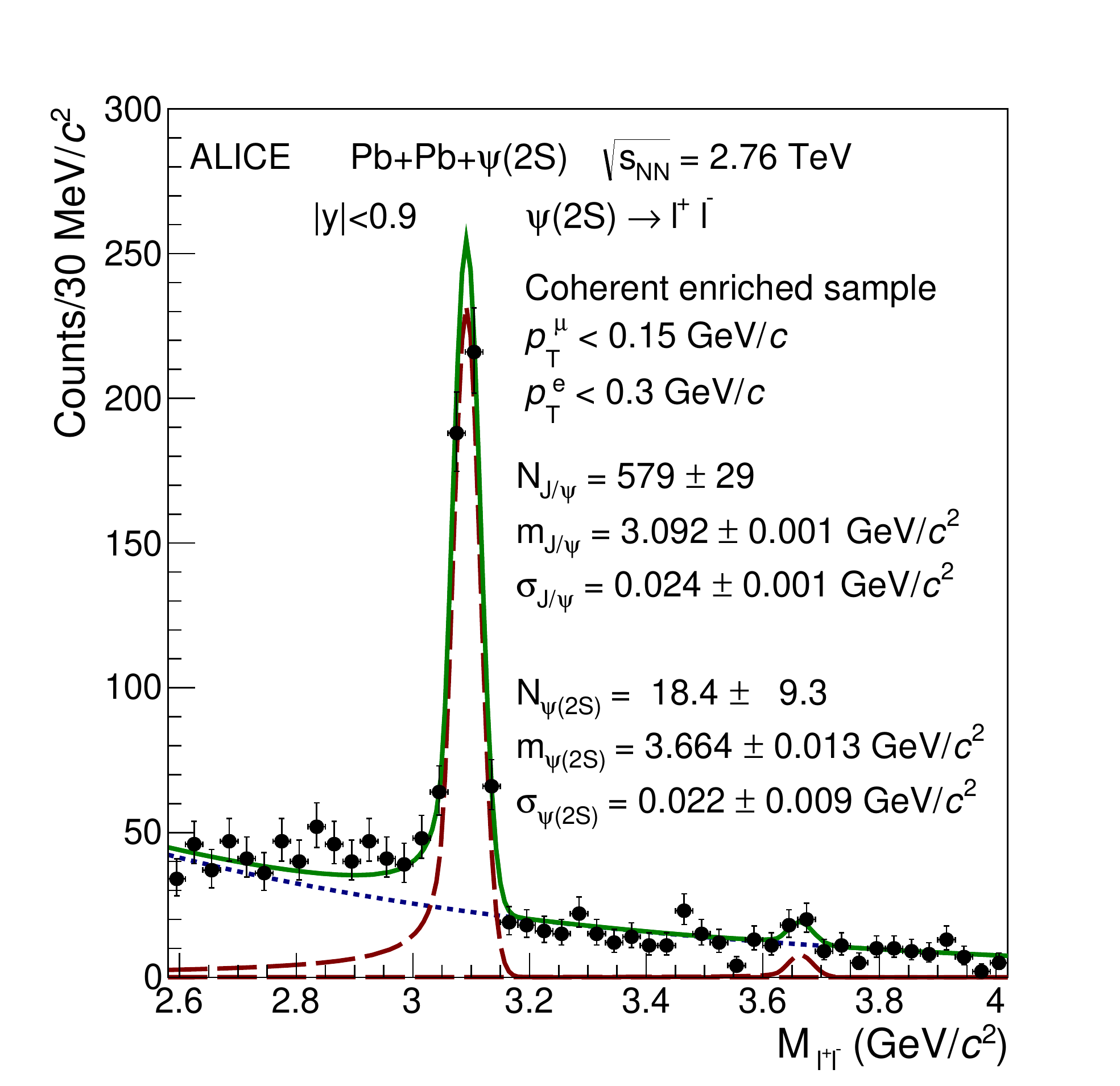}
\includegraphics[width=7cm,keepaspectratio]{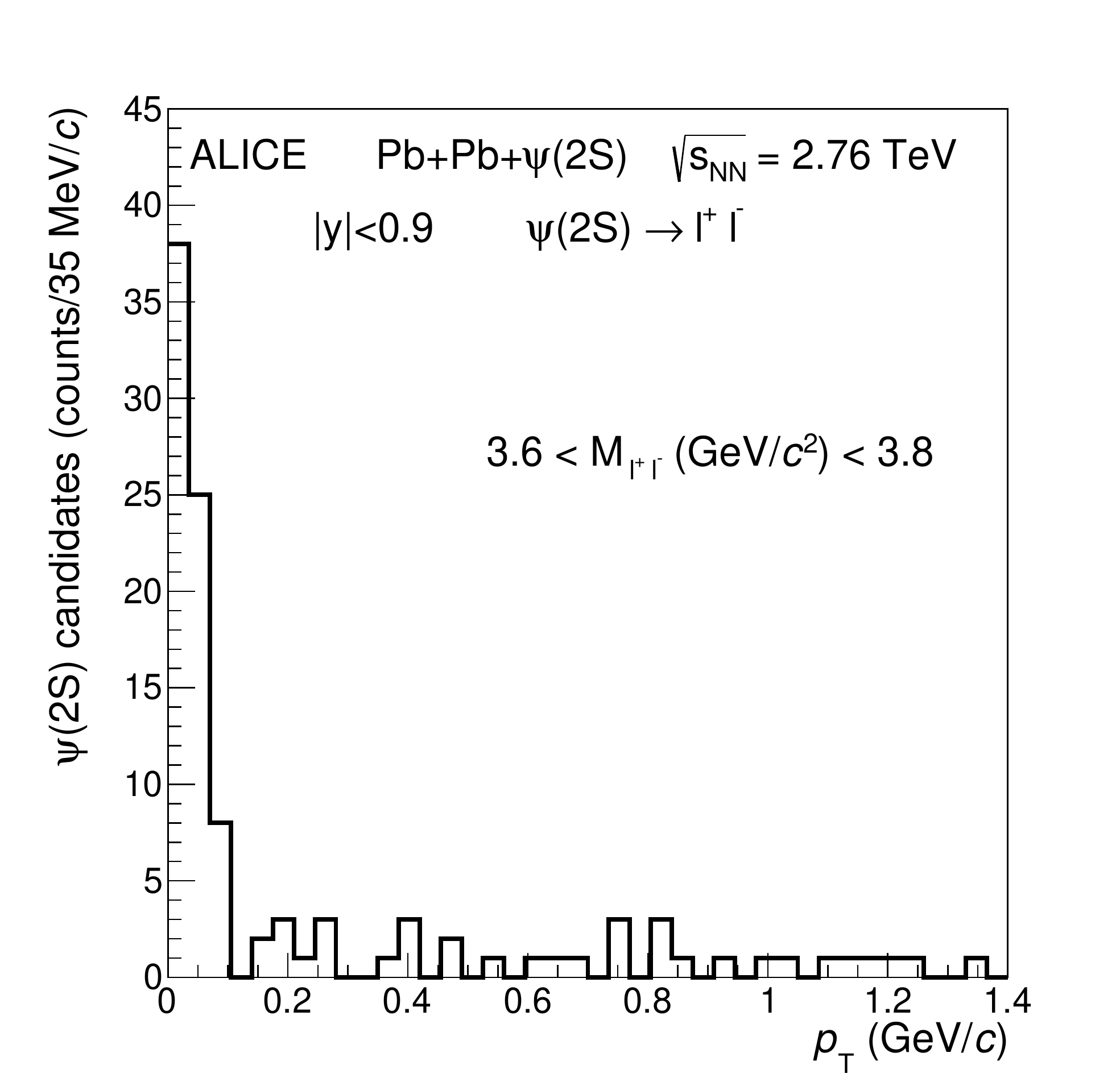}
\includegraphics[width=7cm,keepaspectratio]{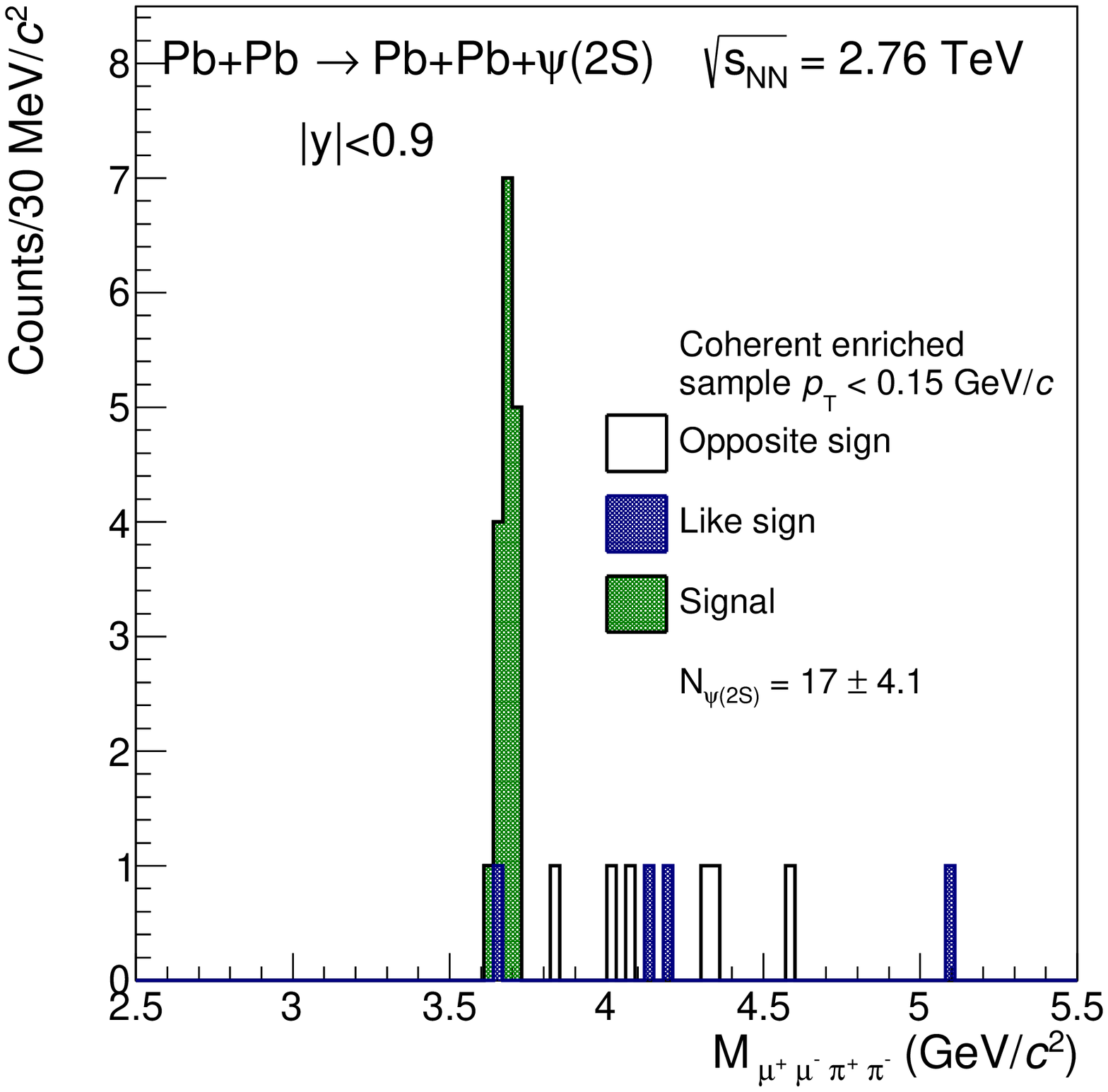}
\includegraphics[width=7cm,keepaspectratio]{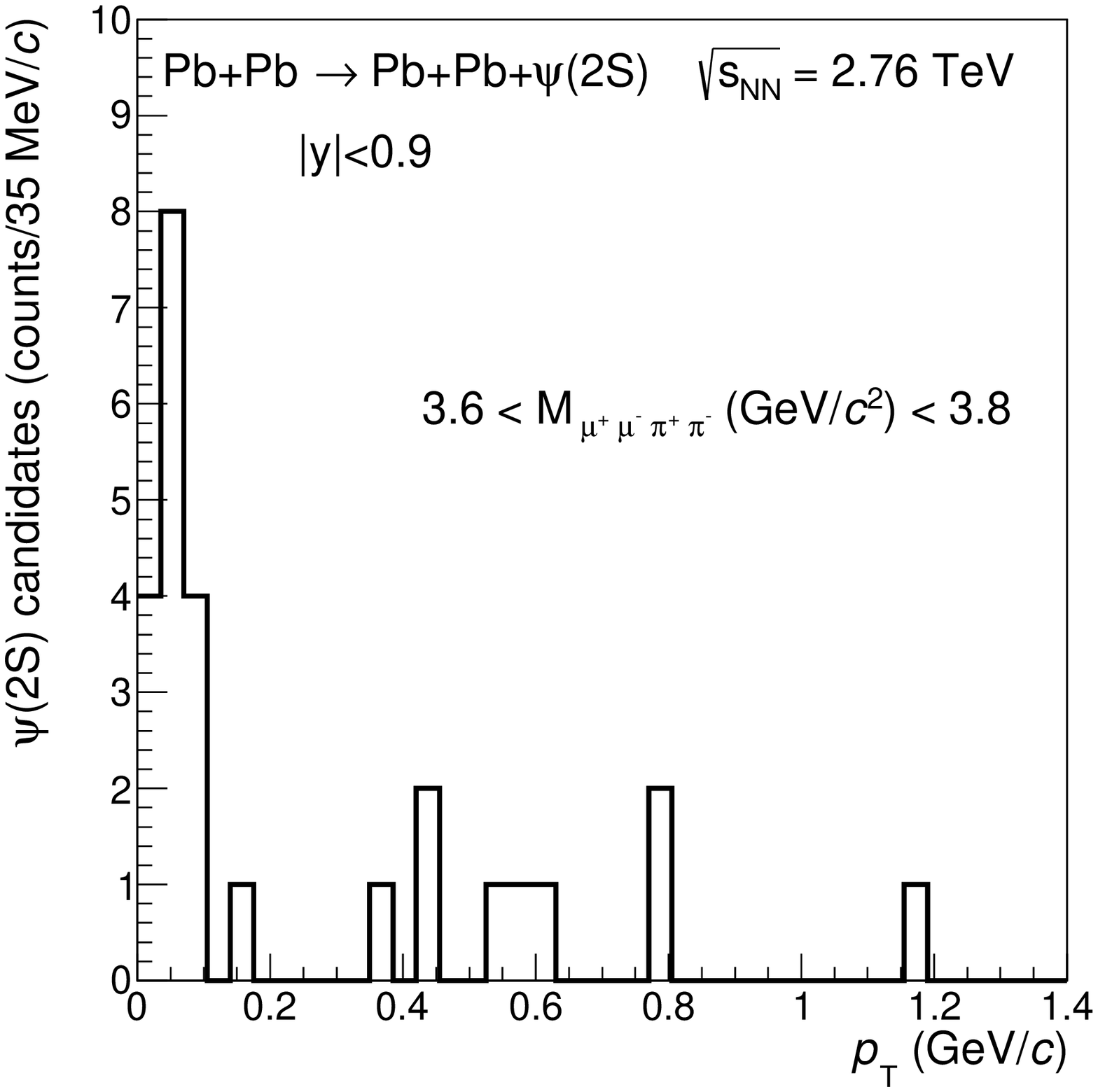}
\includegraphics[width=7cm,keepaspectratio]{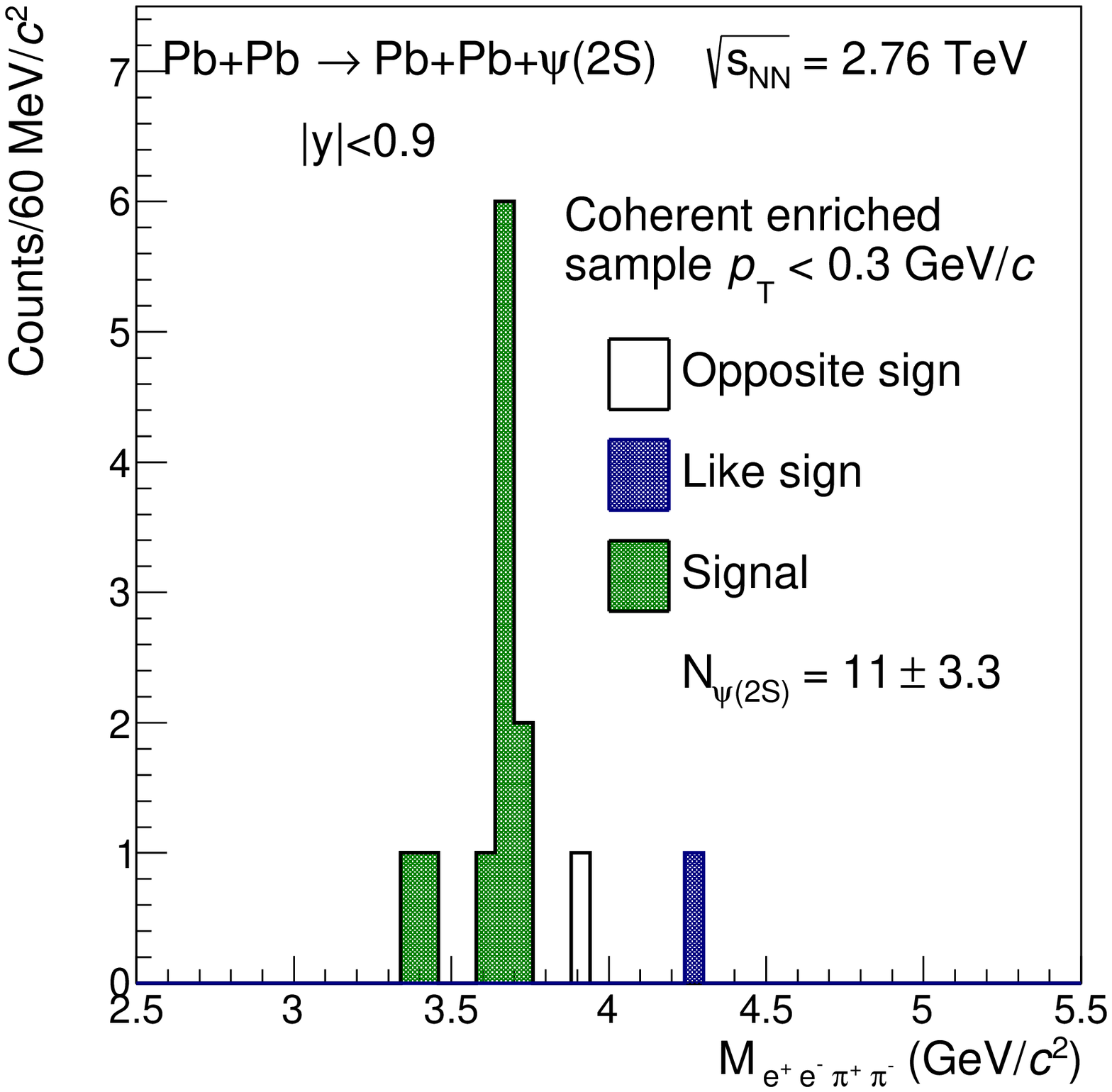}
\includegraphics[width=7cm,keepaspectratio]{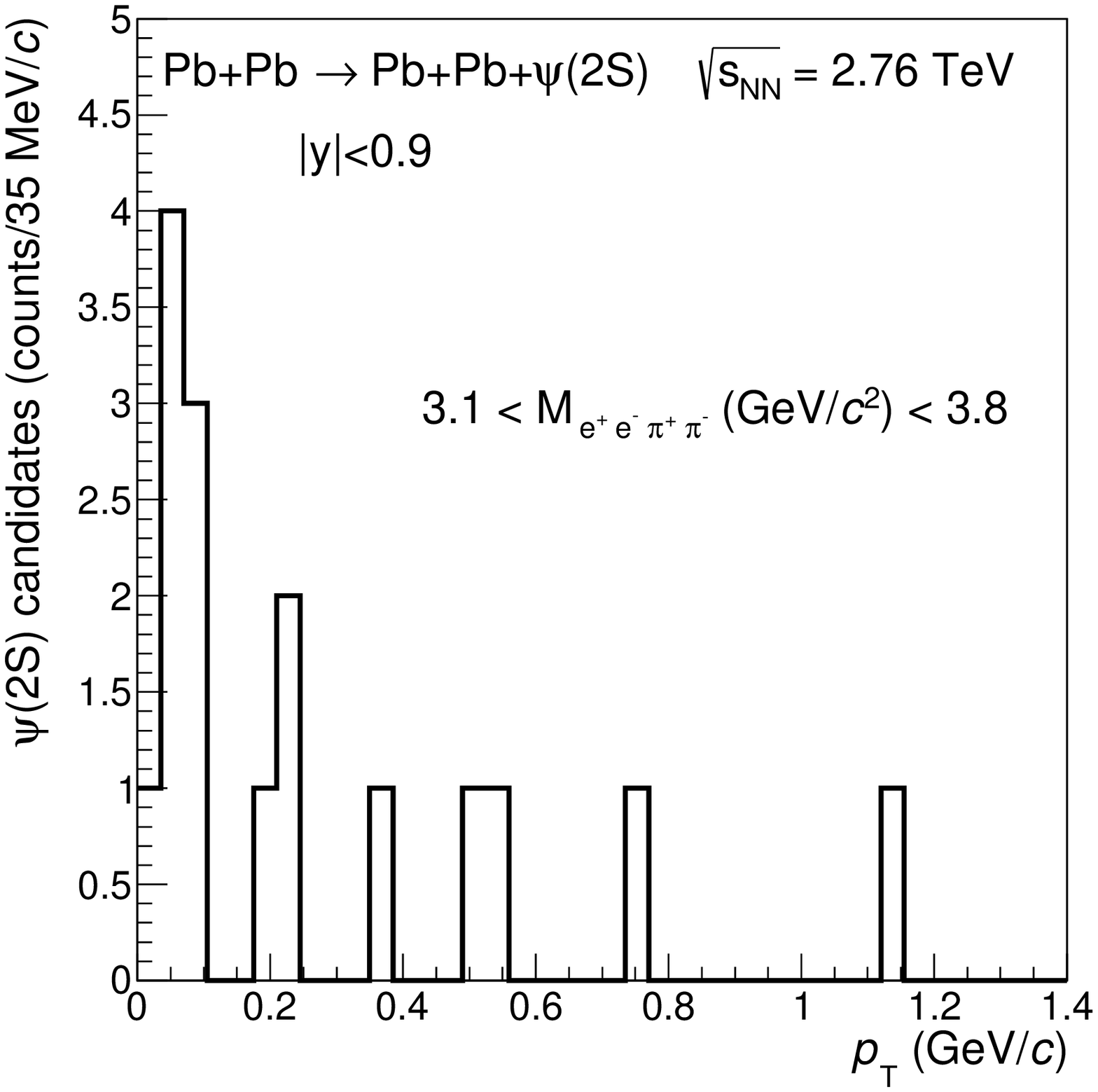}
\caption{Invariant mass~(left) and $p_{T}$ distributions~(right) for ultra-peripheral Pb--Pb collisions at  $\sqrt{s_{\mathrm{NN}} } = 2.76$ TeV and $-0.9<y<0.9$ for 
events satisfying the event selections in Sec.3. 
The channels 
\psip $\rightarrow l^{+}l^{-}$ are shown on the top panel~($l^+l^-=e^{+}e^{-}$ and $\mu^+\mu^-$), the 
\psip $\rightarrow \pi^{+}\pi^{-}\mu^{+}\mu^{-}$ 
channel is shown in the central and the channel \psip $\rightarrow \pi^{+}\pi^{-}e^{+}e^{-}$ in the 
bottom one. The number of event is obtained by the fit (top panel) or by event counting in a selected invariant 
mass region (central and bottom panel), see text.}
\label{fig:yield}
\end{center}
\end{figure}

The optimization of the selection criteria to tag efficiently the \psip~was tailored by using the\\ 
STARLIGHT~\cite{Klein:1999qj} event generator combined with the ALICE detector full simulation. About 950,000 
coherent and incoherent events were simulated for each decay channel.
The event total transverse momentum reconstruction is obtained adding the $p_{T}$ of the two leptons.
The selection of coherent events requires a threshold on the reconstruction of the event total transverse momentum, 
obtained by adding the $p_{T}$ of the two decay leptons. 
Transverse momentum carried away by the bremhstralung photons reflects in a broadening of the event total $p_{T}$. 
Bremhstralung effects are more important for the di-electron decay and the corresponding  $p_{T}$ threshold has to be larger in this case. 
Consequently a $p_{T}$ cut
$p_{T}$$<$0.15 GeV/$c$
for di-muons and $p_{T}$$<$0.3 GeV/$c$ for di-electrons: 98~(77)$\%$ of the coherent signal is retained for 
di-muons and di-electrons respectively.
Figure~\ref{fig:yield} (top panel) shows the invariant mass(left) and the $p_{T}$ distribution(right) for these decay channels.
The $p_{T}$ distributions clearly show a coherent peak at low $p_{T}$. No events are found with a transverse momentum
exceeding 1.5~GeV/$c$, as expected for a negligible hadronic contamination, characterized by a much larger event $p_{T}$.
The number of \psip~candidates are obtained by fitting the invariant mass distribution of both channels to 
an exponential function describing the underlying continuum and to a Crystal Ball function %~\cite{Gaiser:1982t}
to extract the \psip~signal. The Crystal Ball \psip~resonance mass and width were left free, while
the tail parameters ($\alpha$ and $n$) were fixed to the values obtained by Monte Carlo simulation.
The mass (width calculated from the standard deviation) value from the fit is 
$3.664 \pm 0.013$~GeV/$c^2$\ (22$\pm$9 MeV/$c^{2}$)~in good agreement with 
the known value of the \psip~mass and compatible with the absolute calibration 
accuracy of the barrel.
The obtained yield (see Table \ref{tab:CrossSection}) was 
$N^{\rm{yield}}$=(18.4$\pm$9.3).

The product of the acceptance and efficiency correction ($\mathrm{Acc}\times\varepsilon)_{\psip}$ was calculated as 
the ratio of the number of simulated events that satisfy the conditions i) to vi),
to the number of generated events with the \psip~in the rapidity interval $-0.9<y<0.9$. 
Transverse polarization of the \psip~is expected from helicity conservation for a quasi-real photon.
In addition, for the coherent sample, a reconstructed \psip~transverse momentum condition 
$p_{T}$$<$~0.15 GeV/$c$~($p_{T}$$<$~0.3 GeV/$c$) was required for di-muons~(di-electrons) in the final state.
The values for the combined acceptance and efficiency are reported in Table~\ref{tab:CrossSection}.

According to STARLIGHT 
the fraction ($f_{I}$) of incoherent over coherent events in the low~\pt~region is 
4.4$\%$ for di-muons and 16.6$\%$ 
for di-electrons. Another theoretical model, described in~\cite{Guzey:2014kka}, predicts 
a much higher coherent over incoherent cross section ratio, resulting in a
$f_{I}$ prediction 50$\%$ smaller.
Taking the average of
these two predictions,   
(3.3$\pm$1.1)$\%$ for di-muons 
and 
(11.1$\pm$3.4)$\%$ for di-electrons is obtained. The uncertainty was obtained by requiring the used value to agree with the two models
within 1$\sigma$.
The final $f_{I}$ (see Table~\ref{tab:CrossSection}) is the average of the $f_{I}$ for di-electrons and di-muons, weighted 
with the corresponding acceptance and efficiency ({\rm Acc x $\epsilon$}). 
The remaining background~($N_{back}$) was estimated studying the wrong-sign event sample, obtained by applying cuts (i) to (v). 
For di-muon and di-electron channels 
no wrong-sign events were found in the invariant mass range considered and therefore $N_{back}$=0. 

The coherent \psip~yield is obtained using the formula
\begin{equation}
N^{\mathrm{\rm coh}}_{\psip} = \frac{N^{\rm yield}-N^{\rm back}}{1 + f_I} \; ,
\label{NCohJPsi}
\end{equation}
giving $N^{\mathrm{\rm coh}}_{\psip}$=17.5$\pm$9.0.
The coherent \psip~differential cross section can be written as:
\begin{equation}
\label{eq2a2}
\frac
{
\mathrm{d}\sigma_{\psip}^{\mathrm{\rm coh}}
}
{
\mathrm{d}y
}  =
\frac
{N_{\psip}^{\mathrm{\rm coh}}} 
{
(\mathrm{Acc}\times\varepsilon)_{\psip} \cdot {\cal{L}_{\mathrm{int}}} \cdot\Delta y 
\cdot BR(\psip \rightarrow l^{+}l^{-}) 
}
{~~~~~~~~~~~~~~~~~~~~~~~~~~~~~~~~~~~~~~}
\end{equation}
where 
$\acceps$~corresponds to the acceptance and efficiency as discussed above.
$BR(\psip \rightarrow l^{+}l^{-})$ is the branching ratio for \psip~decay into leptons\ ~\cite{Beringer:1900zz},
$\Delta \it{y} =$~1.8 the rapidity bin size, and $\cal{L}_{\mathrm{int}}$
the total integrated luminosity. These values are listed in Table~\ref{tab:CrossSection}.
The systematic uncertainty on the yield for the di-lepton channel 
is obtained by varying the bin size and by replacing the exponential with a polynomial to fit the $\gamma\gamma$ process.
In addition, the Crystal Ball function parameters can be also obtained by fitting a simulated sample made of \psip~and~$\gamma\gamma$ event cocktail 
and then used to fit the coherent-enriched data sample too. 
By applying the different methods reported above, the maximum difference in the obtained yield is 12$\%$: this 
value is used as systematic uncertainty on the yield.
The STARLIGHT model predicts a dependence of the \psip~cross section on the rapidity, giving 
a $\approx$10$\%$ variation over the rapidity range $-0.9<y<0.9$.
In order to evaluate the systematic uncertainty on the acceptance coming from the generator choice,
a flat dependence of $\mathrm{d}\sigma_{\psip}/\mathrm{d}y$ in the interval 
$-0.9<y<0.9$, as predicted by other models, was used.
The relative differences in ($\mathrm{Acc}\times\varepsilon$) between the input shapes was 1.0~$\%$, and are 
taken into account in the systematic uncertainty calculation.
The systematic uncertainty on the tracking efficiency was estimated by comparing, in data 
and in Monte Carlo, the ITS (TPC) hit matching efficiency to tracks reconstructed with 
TPC (ITS) hits only.\\
The trigger efficiency was measured relying on a data sample collected in a dedicated run
triggered by the ZDCs only. 
Events with a topology having the BUPC conditions, given at the beginning of Section 3, were selected.
The resulting trigger efficiency was compared to that obtained by the
Monte Carlo simulation, showing an agreement within $^{+4.0\%}_{-9.0\%}$.\\
The $e$/$\mu$ separation was obtained by using two methods:\\
a) a sharp cut in the scatter plot 
of the first lepton d$E$/d$x$ as a function of the second lepton d$E$/d$x$, 
where all the particles beyond a given threshold 
are considered as electrons;\\
b) using the average of the electron~(muon) d$E$/d$x$ and 
considering as electrons~(muons) the particles within three sigmas from the Bethe-Block expectation. 
The difference between
the two methods was used as an estimate of the systematic uncertainty, giving $\pm 2\%$.\\
The systematic uncertainty related to the application of the V0 offline decision (cut iv) on Sec. 3.1, 
was evaluated repeating the analysis with this cut excluded. This results in a more relaxed event selection, 
increasing the cross section by 6$\%$.\\ 
The integrated luminosity was measured using a trigger for the most central hadronic Pb--Pb collisions.
The cross section for this process was obtained with a van der Meer
scan~\cite{vanderMeer:1968zz}, giving a cross section $\sigma$ = 4.10~$^{+0.22}_{-0.13}$(syst) b~\cite{Abelev:2014ffa}. 
The integrated
luminosity for the BUPC trigger sample, corrected for trigger live time,
was $\cal L_{\rm int}$ = 22.4~$^{+0.9}_{-1.2}~\mu\rm{b}^{-1}$, where the uncertainty 
is the quadratic sum of the cross section uncertainty quoted above and the trigger dead time uncertainty.
An alternative method based on using neutrons detected in the two ZDCs was also used.
The ZDC trigger condition required a signal in at least one of the two calorimeters, thus selecting single
electromagnetic dissociation as well as hadronic interactions.
The cross section for this trigger was also measured with a van der Meer scan~\cite{ALICE:2012aa}.
The integrated luminosity obtained for the BUPC by this method is 
consistent with the one quoted above within 2.5$\%$.
The sources and the values of the systematic uncertainties are listed in Table~\ref{tab:Systematics}.
As a result in the
rapidity interval $\mathrm{-~0.9<{\it y}<0.9}$ a 
cross section $\mathrm{d}\sigma_{\psip}^{\mathrm{\rm coh}}/\mathrm{d}y =\newline
0.76\pm 0.40(\mathrm{stat})^{+0.12}_{-0.13}(\mathrm{syst})$~mb is obtained.
%
%***************** section 3.2 ***************
%
\subsection{\texorpdfstring{The \psip$\rightarrow\pi^+\pi^-$\jpsi,~\jpsi$\rightarrow~l^+l^-(l^+l^-=e^+e^-,\mu^+\mu^-)$ channels}{}}
The analysis criteria used to select these channels are similar to those described in Sec. 3.1, with 
the requirements on the track quality slightly relaxed to keep the efficiency 
at an acceptable level. Such a cut softening was allowed by the smaller 
QED background in four track events, compared to the channels
described in Sec. 3.1. Selection (ii) is modified so that four good tracks with at least 50 TPC clusters each are required.
In addition to cuts i) to vi), the invariant mass of di-muons(di-electrons) 
was required to match that expected by leptons from \jpsi~decay, i.e.      
3.0~$<$~\minvm~$<$~3.2 GeV/$c^{2}$ for di-muons (2.6 $<$~\minve~$<$~3.2 GeV/$c^{2}$ 
for di-electrons). 

The acceptance and the efficiency were estimated with similar techniques.
Due to the coupling to the photon, the \psip is transversely polarized.
According to previous experiments~\cite{Bai:2000pd}, \jpsi~and two pions from \psip~decay are in 
s-wave state and thus the \psip~polarization fully transfers to the \jpsi.
When computing the efficiency and acceptance the \psip~is therefore assumed transversely polarized.
A coherent-enriched sample can be obtained by selecting appropriate
regions on invariant mass and \pt, tuned by using a Monte Carlo simulation, as 
described in Sec. 3.1. The same \pt cuts used in Sec.~3 were applied.
By selecting invariant mass in the interval 
3.6~$<$~\minvm~$<$~3.8 GeV/$c^{2}$ (3.1~$<$~\minve~$<$~3.8 GeV/$c^{2}$) 
for the 
\psip$\rightarrow\pi^+\pi^-\mu^+\mu^-$ 
(\psip$\rightarrow\pi^+\pi^-e^+e^-$) channel, 95$\%$~(87$\%$) of the signal was retained.
The striped area in the invariant mass(left) plots on Fig.~\ref{fig:yield}(central and bottom
panels) shows the \psip~candidates satisfying the \pt~cut for the two channels. To extract the coherent 
\psip~yield, the contribution from incoherent \psip~was subtracted as shown in Eq.(1). 
The background was estimated by looking at 
events with all the possible combination of wrong-sign tracks. 
One event was found 
in the di-muon sample and no events in the di-electron sample.
The fraction of the incoherent sample contaminating the coherent sample was estimated 
as in Sec. 3.1, and was found to be 3.4$\%$ in the
\psip $\rightarrow\pi^+\pi^{-}\mu^{+}\mu^{-}$ channel
and 13.2$\%$ in the 
\psip $\rightarrow\pi^+\pi^- e^+e^-$ channel.
The systematic uncertainty on the yield was obtained by
using an alternative set of cuts. According to the 
kinematics of the \psip$\rightarrow\pi^{+}\pi^{-}\jpsi$ decay channel, pions are 
characterized by a small transverse momentum ($p_{T}<$0.4~GeV/$c$), while the lepton transverse 
momentum exceeds 1.1~GeV/$c$. 
Instead of selecting events where the di-lepton invariant mass is close to that of the \jpsi,
events were selected according to the kinematics of the decay products of the \psip. All the 
other cuts were kept as described in Sec.~3.1. Two alternative selections were considered: (i) a
sample where both leptons have a transverse momentum larger than 1.1~GeV/$c$; and (ii) a 
sample without any decay product with transverse momentum
in the range  0.4$<p_{T}<$1.2~GeV/$c$. The \psip~yield was unchanged for both these selections 
while a small change applies to the acceptance and efficiency in the $\pi^{+}\pi^{-}\jpsi$  
decay, giving a negligible systematic uncertainty.
The relative difference in ($\mathrm{Acc}\times\varepsilon$) between the STARLIGHT rapidity shape and 
a flat rapidity one was 2.0~$\%$ for 
\psip$\rightarrow\pi^+\pi^-\jpsi$ channel, and is
taken into account in the systematic uncertainty calculation.
As a result the obtained 
cross sections in the
rapidity interval $-0.9<y<0.9$ 
are $\mathrm{d}\sigma_{\psip}^{\rm coh}/\mathrm{d}y =
0.81 \pm 0.22(\mathrm{stat})^{+0.09}_{-0.10}(\mathrm{syst})$~mb for the 
\psip$\rightarrow\pi^+\pi^-\jpsi,~\jpsi\rightarrow\mu^+\mu^-$ 
channel and 
$\mathrm{d}\sigma_{\psip}^{\rm coh}/\mathrm{d}y =
0.89 \pm 0.31(\mathrm{stat})^{+0.13}_{-0.12}(\mathrm{syst})$~mb for the 
\psip$\rightarrow\pi^+\pi^-\jpsi,~\jpsi\rightarrow e^+e^-$ 
channel.
%
%
%***************** section 3.3 ***************
%
\subsection{Combining the cross sections}
\begin{figure}[tbh!]
\begin{center}
\includegraphics[width=0.9\linewidth,keepaspectratio]{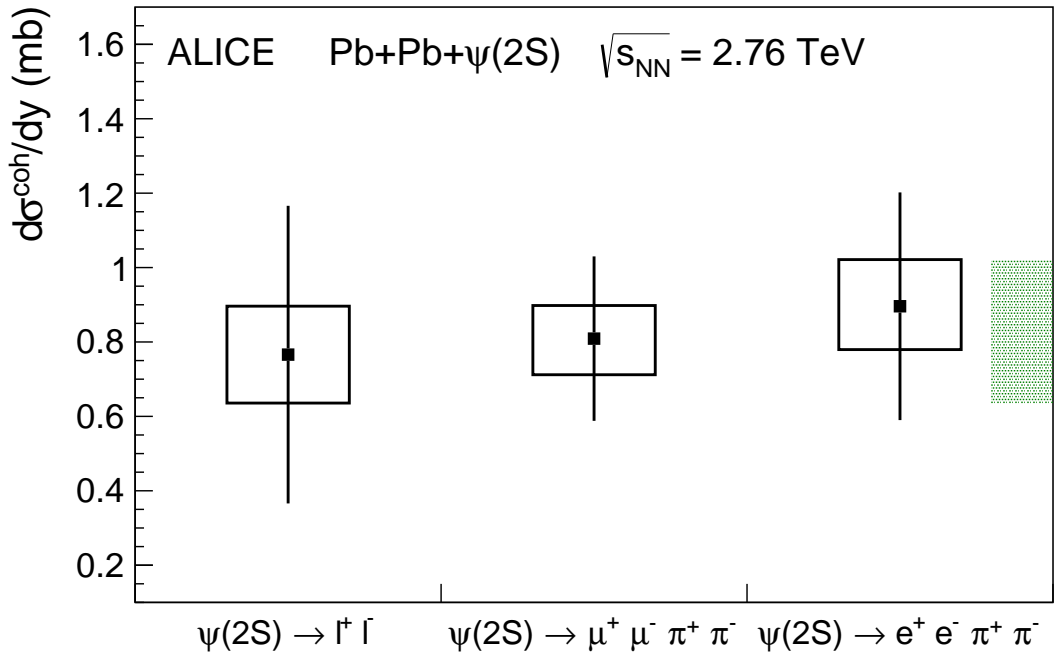}\hspace*{5cm}
\caption{\label{fig:cross} Measured differential cross section of \psip~photo-production in Pb--Pb 
ultra-peripheral 
collisions at \snn=2.76 TeV at $-0.9<y<0.9$ in three different channels. 
The square represents the systematic uncertainties while the bar represents the statistic uncertainty. 
The combined cross section uncertainty (shaded 
area) was obtained using the prescription from reference~\cite{Barlow:2004wg}.}
\end{center}
\end{figure}

The \psip~coherent production cross sections reported in the Sections 3.1 and 
3.2~(Fig.~\ref{fig:cross}) were combined, using the statistical and the uncorrelated systematic uncertainty as a weight.
Finally the correlated systematic uncertainty was added. 
Asymmetric uncertaintys were combined 
according to the prescriptions given in~\cite{Barlow:2004wg}.
The average cross section in the rapidity interval $-0.9<y<0.9$ is 
$\mathrm{d}\sigma_{\psip}^{\rm coh}/\mathrm{d}y =0.83\pm 0.19\big(\mathrm{\rm{stat}+{\rm syst}}\big)$~mb. 

\begin{figure}[tbh!]
\begin{center}
\includegraphics[width=0.9\linewidth,keepaspectratio]{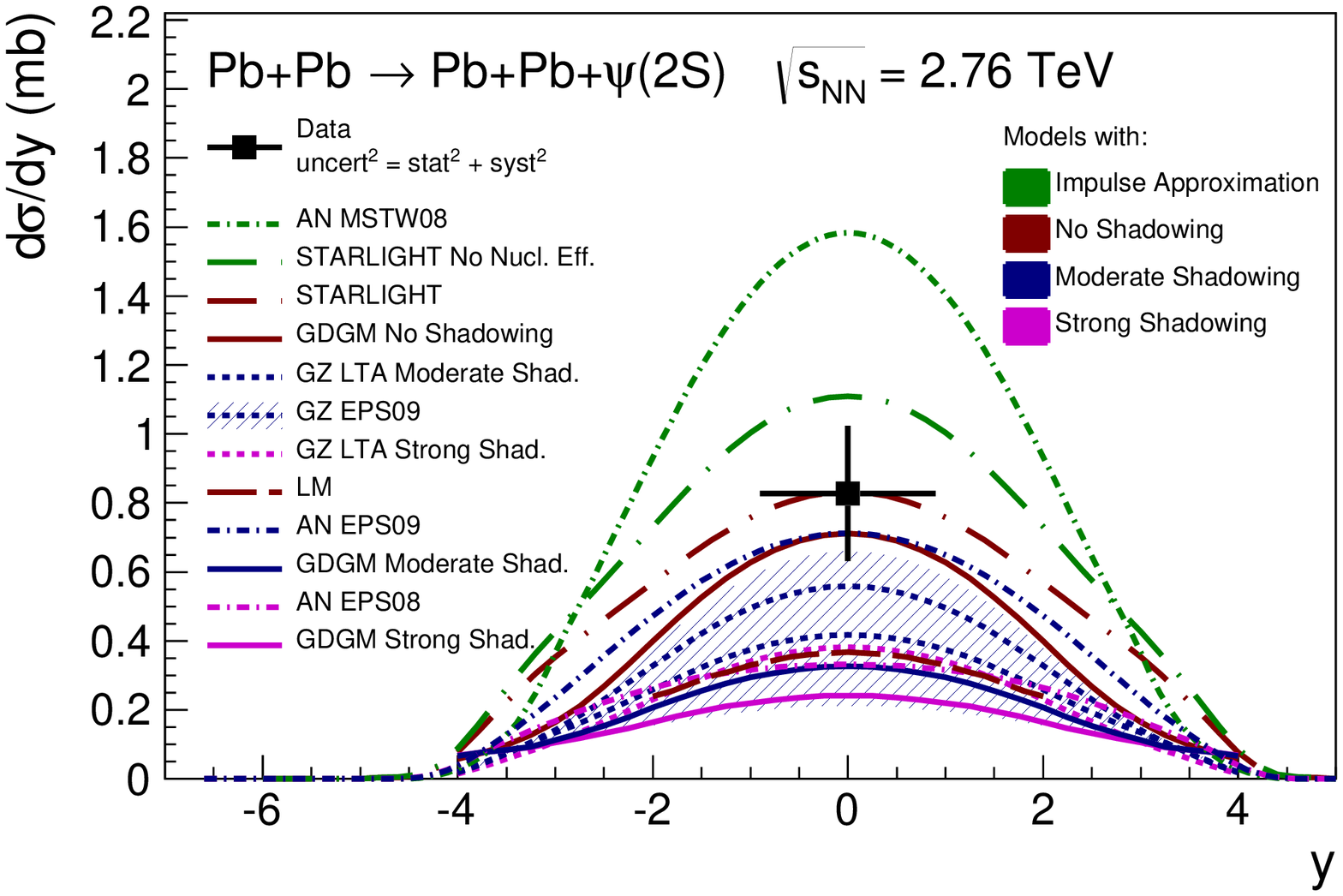}
\caption{\label{fig:abscross} Measured differential cross section of \psip~photo-production
in ultra-peripheral Pb--Pb collisions at \snn=2.76 TeV at $-0.9<y<0.9$.
The uncertainty was obtained using the prescription from reference~\cite{Barlow:2004wg}.
The theoretical calculations described in the text are also shown.}
\end{center}
\end{figure}

\subsection{Coherent production with nuclear break up or nucleus de-excitation followed by neutron emission}
In UPC one or both nuclei may get excited due to the exchange of additional photons. 
This excitation may lead to break up of the nucleus via emission of one or more neutrons. 
The neutron emission was measured by using the ZDC detector, for the events studied in the decay channel, 
$\rm \psi(2S) \rightarrow l^{+} l^{-} \pi^{+} \pi^{-}$. 
We found 20 events (71 $^{+9}_{-11}) \%$ with no neutrons on either side~(0n,0n), 
8 events (29 $^{+11}_{-9}) \%$ with at least one neutron on 
either side~(Xn), 
7 events (25 $^{+10}_{-8}) \%$ with no neutron on one side and at least one neutron on the other one~(0n~Xn)~and 
1 event (4 $^{+8}_{-3}) \%$ with at least one neutron on both sides~(Xn~Xn). 
Uncertainties on the fraction are obtained assuming a binomial distribution.
These fractions are in agreement with predictions by STARLIGHT~\cite{Baltz:2002pp} and RSZ~\cite{Rebyakova:2011vf},
as shown in Table~\ref{tab:NeutronZDC}.

\subsection{The \texorpdfstring{\psip~to \jpsi~cross section ratio}{}}

In order to compare the coherent \psip~cross section to the previously measured \jpsi~cross section~\cite{Abbas:2013oua}, we report on the~ 
\psip/~\jpsi~cross section ratio. Many of the systematic uncertainties of these measurements are correlated and cancel out in the ratio.
Since the analysis relies on the same data sample and on the same trigger, the systematic uncertainties
for the luminosity evaluation, trigger efficiency, and dead time were considered as fully correlated. Several uncertainties, corresponding 
to the same quantity, measured at slightly different energies (corresponding to the different masses), are partially correlated, 
while the uncorrelated part is small. Namely, the systematic uncertainties for 
e/$\mu$ separation and the measurement of the neutron number are strongly correlated and hence can be neglected in the ratio. The systematic uncertainties 
connected to the signal extraction and the branching ratio
are considered 
uncorrelated between the two measurements.
The quadratic sum of these
sources together with the statistic uncertainty was used to combine different channels in both measurements. 
For the combination of asymmetric uncertainties the prescription from reference~\cite{Barlow:2004wg} was used. 
The value of the ratio is 
($\mathrm{d}\sigma_{\psip}^{\mathrm{\rm coh}} /\mathrm{d}y)/(
\mathrm{d}\sigma_{\jpsi}^{\mathrm{\rm coh}} /\mathrm{d}y) =0.34^{+0.08}_{-0.07}(\rm{stat}+{\rm syst})$. 

\section{Discussion}
We have previously measured the coherent photo-production 
cross section for the J/$\psi$ vector meson at mid and forward rapidities~\cite{Abelev:2012ba,Abbas:2013oua}. The results 
showed that the measured cross section was in good agreement with models that 
include a nuclear gluon shadowing consistent with the EPS09 parametrization~\cite{Eskola:2009uj}. 
Models based on the colour dipole model or hadronic interactions 
of the J/$\psi$ with nuclear matter were disfavoured. The $\psi(2S)$ is similar 
to the J/$\psi$ in its composition ($c \overline{c}$) and mass, but it has a 
more complicated wave function as a consequence of it being a 2S rather than 
a 1S state, and has a larger radius. There is a consensus view about the presence 
of a node in the \psip~wavefunction: few authors pointed out that this node 
gives a natural explanation of the \psip~smaller cross section 
compared to the \jpsi~one; in addition it was argued that the node may give strong 
cancellations in the scattering amplitude in $\gamma$-nucleus interactions~\cite{Nemchik:1996cw,Hufner:2000jb}.

In Pb--Pb collisions
the poor knowledge of the \psip~wave function as a function of the transverse quark pair 
separation $d$ makes it difficult to estimate the nuclear matter effects.

\begin{figure}[tbh!]
\begin{center}
\includegraphics[width=0.9\linewidth,keepaspectratio]{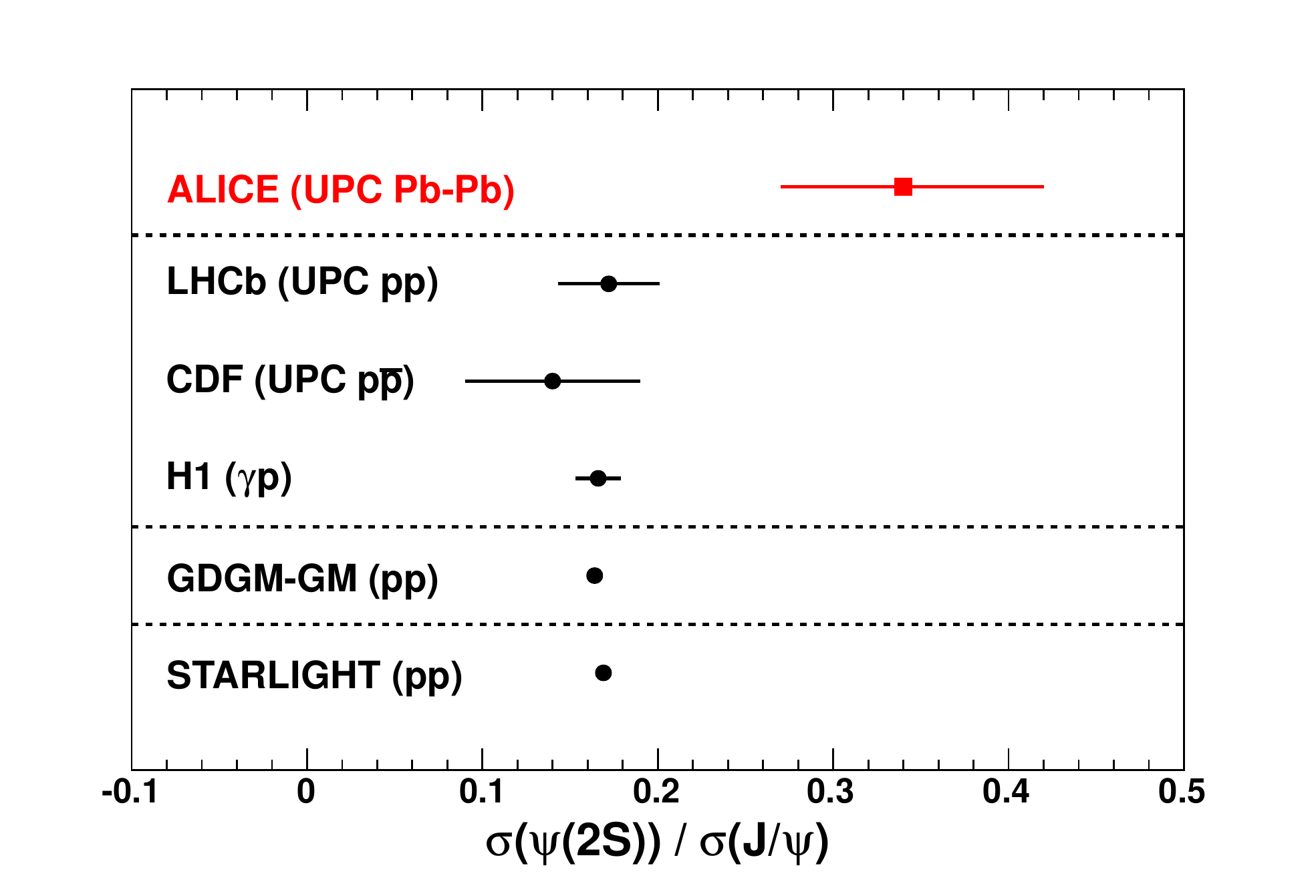}
\caption{\label{fig:ratiopp} Ratio of the \psip~to \jpsi~cross section for pp and $\gamma$p~interactions 
compared to theoretical predictions. The ALICE ratio measured in Pb--Pb collisions is shown as well.
The uncertainty was obtained using the prescription from reference~\cite{Barlow:2004wg}.
}
\end{center}
\end{figure}

There are predictions by five different groups for coherent $\psi(2S)$ production in ultra-peripheral 
Pb--Pb collisions; some of them published several different calculations~(see Fig.\ref{fig:abscross}). 
The model by Adeluyi and Nguyen~(AN) is based on a calculation where 
the \psip~cross section is directly proportional to the gluon distribution 
squared~\cite{Adeluyi:2013tuu}. It is essentially the same model used by Adeluyi and 
Bertulani~\cite{Adeluyi:2012ph} to calculate the coherent J/$\psi$ cross section, which was found 
to be in good agreement with the ALICE data, when coupled to the EPS09 shadowing parametrization. 
The calculations are done for four different 
parameterizations of the nuclear gluon distribution: EPS08~\cite{Eskola:2008}, EPS09~\cite{Eskola:2009uj}, HKN07~\cite{Hirai:2007sx}, and 
MSTW08~\cite{Martin:2009iq}. The last one (MSTW08) corresponds to a scaling of the $\gamma p$ 
cross section neglecting any nuclear effects (impulse approximation). 
It is worth noting they used for the \psip~the same wave function used for the \jpsi.
The model by Gay Ducati, Griep, and Machado~(GDGM)~\cite{Ducati:2013bya} is based on 
the colour dipole model and is similar to the model by Goncalves and Machado for 
coherent J/$\psi$ production~\cite{Goncalves:2011vf}. The latter calculation 
could not reproduce the ALICE coherent \jpsi~measurement. The new calculation has, 
however, been tuned to correctly reproduce the ALICE J/$\psi$ result. 
The model by Lappi and Mantysaari~(LM) is based on the colour dipole model~\cite{Lappi:2013am}. Their prediction for the 
\jpsi~was about a factor of two above the cross section measured by ALICE. 
The current \psip~cross section has been scaled down to compensate for this discrepancy.
The model by Guzey and Zhalov (GZ) is based on the leading approximation of perturbative QCD~\cite{Guzey:2014kka}. They used 
different gluon shadowing predictions, using the dynamical leading twist theory or the EPS09 fit. 
Finally, STARLIGHT uses the Vector Meson Dominance model and a parametrization 
of the existing HERA data to calculate the $\psi(2S)$ cross section from a 
Glauber model assuming only hadronic interactions of the $\psi(2S)$~\cite{Klein:1999qj}. This model 
does not implement nuclear gluon shadowing.

It is worth noting that removing all nuclear effects in STARLIGHT gives a cross section for \jpsi~production
almost identical to the Adeluyi-Bertulani model, if the MSTW08 parametrization is used. The last one corresponds
to a scaling of the $\gamma$-p cross section neglecting any nuclear effects, i.e. considering all nucleons contributing to the 
scattering~(impulse approximation).
Conversely, when applying the same procedure to the \psip~vector meson production, the comparison 
shows that STARLIGHT cross section is 
$\simeq$50$\%$ smaller with respect to the Adeluyi-Nguyen one. 
This result may indicate that the $\gamma$+p$\rightarrow$\psip+p cross section is parametrized in a 
different way in the two models, due to the limited experimental data, making it difficult to tune 
the models. 
For \jpsi,~a wealth of $\gamma +p\rightarrow\jpsi+p$ cross section data has been 
obtained by ZEUS and H1, while the 
process $\gamma +p\rightarrow\psip+p$ was measured by H1 at four different energies only. This makes
it much harder to constrain  
the theoretical cross section to the experimental data. 
Since the effect of gluon shadowing decreases the vector meson production cross section, this  
may explain why the 
\psip~STARLIGHT cross section is close to the AN-EPS09 model, while it is a factor of two larger 
for \jpsi.

The coherent \psip~photo-production cross section is compared to calculations from twelve different models in Fig.~\ref{fig:abscross}.
Since a comprehensive model uncertainty is not provided by the model authors, the comparison with the experimental results 
is quantified by dividing the difference between the value of each model at $y=0$ and the experimental result, by 
the uncertainty of the measurement itself.
The present measurement disfavours the EPS08 
parametrization when implemented in the AN model and the GDGM models with a strong shadowing. 
Similarly the models that neglect any nuclear effect are disfavoured at a level between 1.5 and 3 sigmas.
The systematic uncertainties on the 
cross section parametrization and the experimental statistical uncertainties do not allow 
a preference to be given between the models implementing moderate
nuclear gluon shadowing (as AN-EPS09) and those taking into account Glauber nuclear effects only (as STARLIGHT).

\begin{figure}[tbh!]
\begin{center}
\includegraphics[width=0.9\linewidth,keepaspectratio]{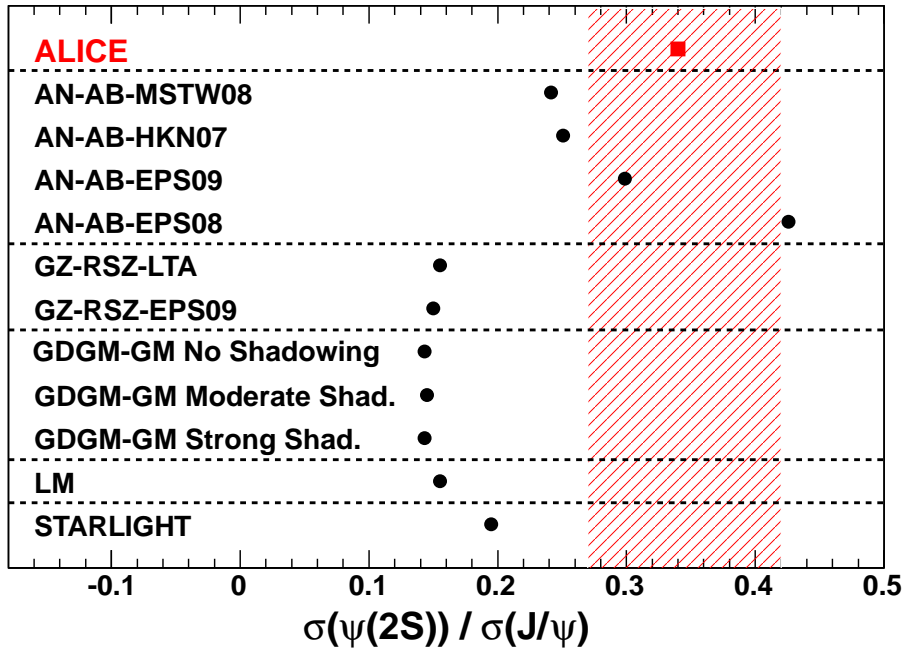}
\caption{\label{fig:ratioAA} Ratio of the \psip~to~\jpsi~cross section measured by ALICE in Pb--Pb collisions.
The uncertainty was obtained using the prescription from reference~\cite{Barlow:2004wg}.
The predictions from different theoretical models are also shown.}
\end{center}
\end{figure}

Figure~\ref{fig:ratiopp} shows 
the \psip~to \jpsi~cross section ratio measured in Pb--Pb collisions by ALICE and 
those obtained in $\rm{p\bar{p}}$ collisions by CDF~\cite{Aaltonen:2009kg}, and in pp collisions by LHCb\cite{Aaij:2014iea}. 
Both STARLIGHT and the GDGM model predict correctly the experimental pp results. The figure also shows 
the ratio measured by H1 in $\gamma$p collisions.
The H1 result is compatible with the pp measurements, while the ALICE point is 2$\sigma$ larger than the average 
of the pp measurements, although still
with sizable uncertainties.
This  
difference may indicate that 
the nuclear effects and/or the gluon shadowing 
modify the \jpsi~and the \psip~production in a different way, since other effects, as the different photon flux, 
due to the larger \psip~mass, could not explain such a difference.

Figure~\ref{fig:ratioAA} shows the comparison of the \psip~to \jpsi~cross section ratio between measurements and predictions in Pb--Pb UPC.
Most models predict a \psip~to \jpsi~cross section ratio in Pb--Pb collisions smaller by 2-2.5~$\sigma$ than the one 
measured by ALICE. It is 
worth noting the same models 
which reproduced correctly the pp ratio, fail in describing the Pb--Pb ratio. It is surprising that the AN model, although it assumes 
a \psip~wave function identical to the \jpsi~one, describes in a satisfactory way this ratio.

\section{Conclusions}
We performed the first measurement of the coherent \psip~photo-production cross section 
in Pb--Pb collisions, obtaining 
$\mathrm{d}\sigma_{\psip}^{\rm coh}/\mathrm{d}y =0.83\pm 0.19\big(\mathrm{\rm{stat}+{\rm syst}}\big)$~mb in the interval $\mathrm{-0.9<y<0.9}$. 
This result disfavours models
considering all nucleons contributing to the scattering and those implementing 
strong shadowing, as EPS08 parametrization.
The ratio of the \psip~to \jpsi~cross section ratio in the rapidity interval $-0.9<y<0.9$ is $0.34^{+0.08}_{-0.07}(\rm{stat}+{\rm syst})$.
Most of the models underpredict this ratio by 2-2.5 ~$\sigma$. 
The current models of the \psip~production in ultra-peripheral collisions 
require further efforts; the data shown in the present analysis may help 
to improve the understanding of this process and to refine the theory behind the exclusive vector meson
photo-production.

% acknowledgements
\newenvironment{acknowledgement}{\relax}{\relax}
\begin{acknowledgement}
\section*{Acknowledgements}
% $Id: acknowledgements.tex 2098 2015-04-24 15:54:16Z loizides $
% Version: Jan 2015

The ALICE Collaboration would like to thank all its engineers and technicians for their invaluable contributions to the construction of the experiment and the CERN accelerator teams for the outstanding performance of the LHC complex.
The ALICE Collaboration gratefully acknowledges the resources and support provided by all Grid centres and the Worldwide LHC Computing Grid (WLCG) collaboration.
The ALICE Collaboration acknowledges the following funding agencies for their support in building and
running the ALICE detector:
State Committee of Science,  World Federation of Scientists (WFS)
and Swiss Fonds Kidagan, Armenia,
Conselho Nacional de Desenvolvimento Cient\'{\i}fico e Tecnol\'{o}gico (CNPq), Financiadora de Estudos e Projetos (FINEP),
Funda\c{c}\~{a}o de Amparo \`{a} Pesquisa do Estado de S\~{a}o Paulo (FAPESP);
National Natural Science Foundation of China (NSFC), the Chinese Ministry of Education (CMOE)
and the Ministry of Science and Technology of China (MSTC);
Ministry of Education and Youth of the Czech Republic;
Danish Natural Science Research Council, the Carlsberg Foundation and the Danish National Research Foundation;
The European Research Council under the European Community's Seventh Framework Programme;
Helsinki Institute of Physics and the Academy of Finland;
French CNRS-IN2P3, the `Region Pays de Loire', `Region Alsace', `Region Auvergne' and CEA, France;
German Bundesministerium fur Bildung, Wissenschaft, Forschung und Technologie (BMBF) and the Helmholtz Association;
General Secretariat for Research and Technology, Ministry of
Development, Greece;
Hungarian Orszagos Tudomanyos Kutatasi Alappgrammok (OTKA) and National Office for Research and Technology (NKTH);
Department of Atomic Energy and Department of Science and Technology of the Government of India;
Istituto Nazionale di Fisica Nucleare (INFN) and Centro Fermi -
Museo Storico della Fisica e Centro Studi e Ricerche "Enrico
Fermi", Italy;
MEXT Grant-in-Aid for Specially Promoted Research, Ja\-pan;
Joint Institute for Nuclear Research, Dubna;
National Research Foundation of Korea (NRF);
Consejo Nacional de Cienca y Tecnologia (CONACYT), Direccion General de Asuntos del Personal Academico(DGAPA), M\'{e}xico, Amerique Latine Formation academique - European Commission~(ALFA-EC) and the EPLANET Program~(European Particle Physics Latin American Network);
Stichting voor Fundamenteel Onderzoek der Materie (FOM) and the Nederlandse Organisatie voor Wetenschappelijk Onderzoek (NWO), Netherlands;
Research Council of Norway (NFR);
National Science Centre, Poland;
Ministry of National Education/Institute for Atomic Physics and National Council of Scientific Research in Higher Education~(CNCSI-UEFISCDI), Romania;
Ministry of Education and Science of Russian Federation, Russian
Academy of Sciences, Russian Federal Agency of Atomic Energy,
Russian Federal Agency for Science and Innovations and The Russian
Foundation for Basic Research;
Ministry of Education of Slovakia;
Department of Science and Technology, South Africa;
Centro de Investigaciones Energeticas, Medioambientales y Tecnologicas (CIEMAT), E-Infrastructure shared between Europe and Latin America (EELA), Ministerio de Econom\'{i}a y Competitividad (MINECO) of Spain, Xunta de Galicia (Conseller\'{\i}a de Educaci\'{o}n),
Centro de Aplicaciones Tecnológicas y Desarrollo Nuclear (CEA\-DEN), Cubaenerg\'{\i}a, Cuba, and IAEA (International Atomic Energy Agency);
Swedish Research Council (VR) and Knut $\&$ Alice Wallenberg
Foundation (KAW);
Ukraine Ministry of Education and Science;
United Kingdom Science and Technology Facilities Council (STFC);
The United States Department of Energy, the United States National
Science Foundation, the State of Texas, and the State of Ohio;
Ministry of Science, Education and Sports of Croatia and  Unity through Knowledge Fund, Croatia.
Council of Scientific and Industrial Research (CSIR), New Delhi, India
   done by webmaster team
\end{acknowledgement}

\bibliographystyle{utphys}
\bibliography{psi2s_cernprint}{}

% appendix with author list
\newpage
\appendix

\section{The ALICE Collaboration}
\label{app:collab}

% Collaboration: CERN-LHC-ALICE
% Generation Date is 2015/Mar/13

% How to use:
%%%%%%%%% appendix with author list
%\appendix
%\section{The ALICE Collaboration}
%\label{app:collab}
%\input{authors-list.tex}  %%%%%%% get the latest version before submitting

\begingroup
\small
\begin{flushleft}
J.~Adam\Irefn{org39}\And
D.~Adamov\'{a}\Irefn{org82}\And
M.M.~Aggarwal\Irefn{org86}\And
G.~Aglieri Rinella\Irefn{org36}\And
M.~Agnello\Irefn{org110}\And
N.~Agrawal\Irefn{org47}\And
Z.~Ahammed\Irefn{org130}\And
S.U.~Ahn\Irefn{org67}\And
I.~Aimo\Irefn{org93}\textsuperscript{,}\Irefn{org110}\And
S.~Aiola\Irefn{org135}
% ...
\renewcommand\labelenumi{\textsuperscript{\theenumi}~}

\section*{Affiliation notes}
\renewcommand\theenumi{\roman{enumi}}
\begin{Authlist}
\item \Adef{0}Deceased
\item \Adef{idp5850416}{Also at: University of Kansas, Lawrence, Kansas, United States}
\end{Authlist}

\section*{Collaboration Institutes}
\renewcommand\theenumi{\arabic{enumi}~}
\begin{Authlist}

\item \Idef{org1}A.I. Alikhanyan National Science Laboratory (Yerevan Physics Institute) Foundation, Yerevan, Armenia
\item \Idef{org2}Benem\'{e}rita Universidad Aut\'{o}noma de Puebla, Puebla, Mexico
\item \Idef{org3}Bogolyubov Institute for Theoretical Physics, Kiev, Ukraine
\item \Idef{org4}Bose Institute, Department of Physics and Centre for Astroparticle Physics and Space Science (CAPSS), Kolkata, India
\item \Idef{org5}Budker Institute for Nuclear Physics, Novosibirsk, Russia
\item \Idef{org6}California Polytechnic State University, San Luis Obispo, California, United States
\item \Idef{org7}Central China Normal University, Wuhan, China
\item \Idef{org8}Centre de Calcul de l'IN2P3, Villeurbanne, France
\item \Idef{org9}Centro de Aplicaciones Tecnol\'{o}gicas y Desarrollo Nuclear (CEADEN), Havana, Cuba
\item \Idef{org10}Centro de Investigaciones Energ\'{e}ticas Medioambientales y Tecnol\'{o}gicas (CIEMAT), Madrid, Spain
\item \Idef{org11}Centro de Investigaci\'{o}n y de Estudios Avanzados (CINVESTAV), Mexico City and M\'{e}rida, Mexico
\item \Idef{org12}Centro Fermi - Museo Storico della Fisica e Centro Studi e Ricerche ``Enrico Fermi'', Rome, Italy
\item \Idef{org13}Chicago State University, Chicago, Illinois, USA
\item \Idef{org14}China Institute of Atomic Energy, Beijing, China
\item \Idef{org15}Commissariat \`{a} l'Energie Atomique, IRFU, Saclay, France
\item \Idef{org16}COMSATS Institute of Information Technology (CIIT), Islamabad, Pakistan
\item \Idef{org17}Departamento de F\'{\i}sica de Part\'{\i}culas and IGFAE, Universidad de Santiago de Compostela, Santiago de Compostela, Spain
\item \Idef{org18}Department of Physics and Technology, University of Bergen, Bergen, Norway
\item \Idef{org19}Department of Physics, Aligarh Muslim University, Aligarh, India
\item \Idef{org20}Department of Physics, Ohio State University, Columbus, Ohio, United States
\item \Idef{org21}Department of Physics, Sejong University, Seoul, South Korea
\item \Idef{org22}Department of Physics, University of Oslo, Oslo, Norway
\item \Idef{org23}Dipartimento di Elettrotecnica ed Elettronica del Politecnico, Bari, Italy
\item \Idef{org24}Dipartimento di Fisica dell'Universit\`{a} 'La Sapienza' and Sezione INFN Rome, Italy
\item \Idef{org25}Dipartimento di Fisica dell'Universit\`{a} and Sezione INFN, Cagliari, Italy
\item \Idef{org26}Dipartimento di Fisica dell'Universit\`{a} and Sezione INFN, Trieste, Italy
\item \Idef{org27}Dipartimento di Fisica dell'Universit\`{a} and Sezione INFN, Turin, Italy
\item \Idef{org28}Dipartimento di Fisica e Astronomia dell'Universit\`{a} and Sezione INFN, Bologna, Italy
\item \Idef{org29}Dipartimento di Fisica e Astronomia dell'Universit\`{a} and Sezione INFN, Catania, Italy
\item \Idef{org30}Dipartimento di Fisica e Astronomia dell'Universit\`{a} and Sezione INFN, Padova, Italy
\item \Idef{org31}Dipartimento di Fisica `E.R.~Caianiello' dell'Universit\`{a} and Gruppo Collegato INFN, Salerno, Italy
\item \Idef{org32}Dipartimento di Scienze e Innovazione Tecnologica dell'Universit\`{a} del  Piemonte Orientale and Gruppo Collegato INFN, Alessandria, Italy
\item \Idef{org33}Dipartimento Interateneo di Fisica `M.~Merlin' and Sezione INFN, Bari, Italy
\item \Idef{org34}Division of Experimental High Energy Physics, University of Lund, Lund, Sweden
\item \Idef{org35}Eberhard Karls Universit\"{a}t T\"{u}bingen, T\"{u}bingen, Germany
\item \Idef{org36}European Organization for Nuclear Research (CERN), Geneva, Switzerland
\item \Idef{org37}Faculty of Engineering, Bergen University College, Bergen, Norway
\item \Idef{org38}Faculty of Mathematics, Physics and Informatics, Comenius University, Bratislava, Slovakia
\item \Idef{org39}Faculty of Nuclear Sciences and Physical Engineering, Czech Technical University in Prague, Prague, Czech Republic
\item \Idef{org40}Faculty of Science, P.J.~\v{S}af\'{a}rik University, Ko\v{s}ice, Slovakia
\item \Idef{org41}Faculty of Technology, Buskerud and Vestfold University College, Vestfold, Norway
\item \Idef{org42}Frankfurt Institute for Advanced Studies, Johann Wolfgang Goethe-Universit\"{a}t Frankfurt, Frankfurt, Germany
\item \Idef{org43}Gangneung-Wonju National University, Gangneung, South Korea
\item \Idef{org44}Gauhati University, Department of Physics, Guwahati, India
\item \Idef{org45}Helsinki Institute of Physics (HIP), Helsinki, Finland
\item \Idef{org46}Hiroshima University, Hiroshima, Japan
\item \Idef{org47}Indian Institute of Technology Bombay (IIT), Mumbai, India
\item \Idef{org48}Indian Institute of Technology Indore, Indore (IITI), India
\item \Idef{org49}Inha University, Incheon, South Korea
\item \Idef{org50}Institut de Physique Nucl\'eaire d'Orsay (IPNO), Universit\'e Paris-Sud, CNRS-IN2P3, Orsay, France
\item \Idef{org51}Institut f\"{u}r Informatik, Johann Wolfgang Goethe-Universit\"{a}t Frankfurt, Frankfurt, Germany
\item \Idef{org52}Institut f\"{u}r Kernphysik, Johann Wolfgang Goethe-Universit\"{a}t Frankfurt, Frankfurt, Germany
\item \Idef{org53}Institut f\"{u}r Kernphysik, Westf\"{a}lische Wilhelms-Universit\"{a}t M\"{u}nster, M\"{u}nster, Germany
\item \Idef{org54}Institut Pluridisciplinaire Hubert Curien (IPHC), Universit\'{e} de Strasbourg, CNRS-IN2P3, Strasbourg, France
\item \Idef{org55}Institute for Nuclear Research, Academy of Sciences, Moscow, Russia
\item \Idef{org56}Institute for Subatomic Physics of Utrecht University, Utrecht, Netherlands
\item \Idef{org57}Institute for Theoretical and Experimental Physics, Moscow, Russia
\item \Idef{org58}Institute of Experimental Physics, Slovak Academy of Sciences, Ko\v{s}ice, Slovakia
\item \Idef{org59}Institute of Physics, Academy of Sciences of the Czech Republic, Prague, Czech Republic
\item \Idef{org60}Institute of Physics, Bhubaneswar, India
\item \Idef{org61}Institute of Space Science (ISS), Bucharest, Romania
\item \Idef{org62}Instituto de Ciencias Nucleares, Universidad Nacional Aut\'{o}noma de M\'{e}xico, Mexico City, Mexico
\item \Idef{org63}Instituto de F\'{\i}sica, Universidad Nacional Aut\'{o}noma de M\'{e}xico, Mexico City, Mexico
\item \Idef{org64}iThemba LABS, National Research Foundation, Somerset West, South Africa
\item \Idef{org65}Joint Institute for Nuclear Research (JINR), Dubna, Russia
\item \Idef{org66}Konkuk University, Seoul, South Korea
\item \Idef{org67}Korea Institute of Science and Technology Information, Daejeon, South Korea
\item \Idef{org68}KTO Karatay University, Konya, Turkey
\item \Idef{org69}Laboratoire de Physique Corpusculaire (LPC), Clermont Universit\'{e}, Universit\'{e} Blaise Pascal, CNRS--IN2P3, Clermont-Ferrand, France
\item \Idef{org70}Laboratoire de Physique Subatomique et de Cosmologie, Universit\'{e} Grenoble-Alpes, CNRS-IN2P3, Grenoble, France
\item \Idef{org71}Laboratori Nazionali di Frascati, INFN, Frascati, Italy
\item \Idef{org72}Laboratori Nazionali di Legnaro, INFN, Legnaro, Italy
\item \Idef{org73}Lawrence Berkeley National Laboratory, Berkeley, California, United States
\item \Idef{org74}Lawrence Livermore National Laboratory, Livermore, California, United States
\item \Idef{org75}Moscow Engineering Physics Institute, Moscow, Russia
\item \Idef{org76}National Centre for Nuclear Studies, Warsaw, Poland
\item \Idef{org77}National Institute for Physics and Nuclear Engineering, Bucharest, Romania
\item \Idef{org78}National Institute of Science Education and Research, Bhubaneswar, India
\item \Idef{org79}Niels Bohr Institute, University of Copenhagen, Copenhagen, Denmark
\item \Idef{org80}Nikhef, National Institute for Subatomic Physics, Amsterdam, Netherlands
\item \Idef{org81}Nuclear Physics Group, STFC Daresbury Laboratory, Daresbury, United Kingdom
\item \Idef{org82}Nuclear Physics Institute, Academy of Sciences of the Czech Republic, \v{R}e\v{z} u Prahy, Czech Republic
\item \Idef{org83}Oak Ridge National Laboratory, Oak Ridge, Tennessee, United States
\item \Idef{org84}Petersburg Nuclear Physics Institute, Gatchina, Russia
\item \Idef{org85}Physics Department, Creighton University, Omaha, Nebraska, United States
\item \Idef{org86}Physics Department, Panjab University, Chandigarh, India
\item \Idef{org87}Physics Department, University of Athens, Athens, Greece
\item \Idef{org88}Physics Department, University of Cape Town, Cape Town, South Africa
\item \Idef{org89}Physics Department, University of Jammu, Jammu, India
\item \Idef{org90}Physics Department, University of Rajasthan, Jaipur, India
\item \Idef{org91}Physik Department, Technische Universit\"{a}t M\"{u}nchen, Munich, Germany
\item \Idef{org92}Physikalisches Institut, Ruprecht-Karls-Universit\"{a}t Heidelberg, Heidelberg, Germany
\item \Idef{org93}Politecnico di Torino, Turin, Italy
\item \Idef{org94}Purdue University, West Lafayette, Indiana, United States
\item \Idef{org95}Pusan National University, Pusan, South Korea
\item \Idef{org96}Research Division and ExtreMe Matter Institute EMMI, GSI Helmholtzzentrum f\"ur Schwerionenforschung, Darmstadt, Germany
\item \Idef{org97}Rudjer Bo\v{s}kovi\'{c} Institute, Zagreb, Croatia
\item \Idef{org98}Russian Federal Nuclear Center (VNIIEF), Sarov, Russia
\item \Idef{org99}Russian Research Centre Kurchatov Institute, Moscow, Russia
\item \Idef{org100}Saha Institute of Nuclear Physics, Kolkata, India
\item \Idef{org101}School of Physics and Astronomy, University of Birmingham, Birmingham, United Kingdom
\item \Idef{org102}Secci\'{o}n F\'{\i}sica, Departamento de Ciencias, Pontificia Universidad Cat\'{o}lica del Per\'{u}, Lima, Peru
\item \Idef{org103}Sezione INFN, Bari, Italy
\item \Idef{org104}Sezione INFN, Bologna, Italy
\item \Idef{org105}Sezione INFN, Cagliari, Italy
\item \Idef{org106}Sezione INFN, Catania, Italy
\item \Idef{org107}Sezione INFN, Padova, Italy
\item \Idef{org108}Sezione INFN, Rome, Italy
\item \Idef{org109}Sezione INFN, Trieste, Italy
\item \Idef{org110}Sezione INFN, Turin, Italy
\item \Idef{org111}SSC IHEP of NRC Kurchatov institute, Protvino, Russia
\item \Idef{org112}SUBATECH, Ecole des Mines de Nantes, Universit\'{e} de Nantes, CNRS-IN2P3, Nantes, France
\item \Idef{org113}Suranaree University of Technology, Nakhon Ratchasima, Thailand
\item \Idef{org114}Technical University of Split FESB, Split, Croatia
\item \Idef{org115}The Henryk Niewodniczanski Institute of Nuclear Physics, Polish Academy of Sciences, Cracow, Poland
\item \Idef{org116}The University of Texas at Austin, Physics Department, Austin, Texas, USA
\item \Idef{org117}Universidad Aut\'{o}noma de Sinaloa, Culiac\'{a}n, Mexico
\item \Idef{org118}Universidade de S\~{a}o Paulo (USP), S\~{a}o Paulo, Brazil
\item \Idef{org119}Universidade Estadual de Campinas (UNICAMP), Campinas, Brazil
\item \Idef{org120}University of Houston, Houston, Texas, United States
\item \Idef{org121}University of Jyv\"{a}skyl\"{a}, Jyv\"{a}skyl\"{a}, Finland
\item \Idef{org122}University of Liverpool, Liverpool, United Kingdom
\item \Idef{org123}University of Tennessee, Knoxville, Tennessee, United States
\item \Idef{org124}University of the Witwatersrand, Johannesburg, South Africa
\item \Idef{org125}University of Tokyo, Tokyo, Japan
\item \Idef{org126}University of Tsukuba, Tsukuba, Japan
\item \Idef{org127}University of Zagreb, Zagreb, Croatia
\item \Idef{org128}Universit\'{e} de Lyon, Universit\'{e} Lyon 1, CNRS/IN2P3, IPN-Lyon, Villeurbanne, France
\item \Idef{org129}V.~Fock Institute for Physics, St. Petersburg State University, St. Petersburg, Russia
\item \Idef{org130}Variable Energy Cyclotron Centre, Kolkata, India
\item \Idef{org131}Vin\v{c}a Institute of Nuclear Sciences, Belgrade, Serbia
\item \Idef{org132}Warsaw University of Technology, Warsaw, Poland
\item \Idef{org133}Wayne State University, Detroit, Michigan, United States
\item \Idef{org134}Wigner Research Centre for Physics, Hungarian Academy of Sciences, Budapest, Hungary
\item \Idef{org135}Yale University, New Haven, Connecticut, United States
\item \Idef{org136}Yonsei University, Seoul, South Korea
\item \Idef{org137}Zentrum f\"{u}r Technologietransfer und Telekommunikation (ZTT), Fachhochschule Worms, Worms, Germany
\end{Authlist}
\endgroup

\end{document}